\documentclass[aps,prl]{revtex4}

\usepackage{graphicx,amssymb,amsmath}

\usepackage{color}

\newcommand{\be}{\begin{equation}}
\newcommand{\ee}{\end{equation}}

\newcommand{\bea}{\begin{eqnarray}}
\newcommand{\eea}{\end{eqnarray}}

\begin{document}

\title{Boundary fluctuation dynamics of a phase-separated domain in planar geometry}

\author{Nicolas Destainville, Nelly Coulonges}
\affiliation{Laboratoire de Physique Th\'eorique, Universit\'e de Toulouse, CNRS, UPS, France}

\date{\today}

\begin{abstract}
Using theories of phase ordering kinetics and of renormalization group, we derive analytically the relaxation times of the long wave-length fluctuations of a phase-separated domain boundary in the vicinity of (and below) the critical temperature, in the planar Ising universality class. For a conserved order parameter, the relaxation time grows like $\Lambda^3$ at wave-length $\Lambda$ and can be expressed in terms of parameters relevant at the microscopic scale: lattice spacing, bulk diffusion coefficient of the minority phase, and  temperature. These  results are supported by numerical simulations of 2D Ising models, enabling in addition to calculate the non-universal numerical prefactor. We discuss the applications of these findings to the determination of the real time-scale associated with elementary Monte Carlo moves from the measurement of  long wave-length relaxation times on experimental systems or Molecular Dynamics simulations. \end{abstract}

\maketitle

Phase separation phenomena are ubiquitous in Nature, and many of them are observed in two dimensions, on surfaces, interfaces or membranes~\cite{Herbut,Seul95}. When they belong to the Ising-universality class~\cite{Honerkamp08,Honerkamp09}, they are often numerically tackled with the help of this Ising model, on a square or a triangular lattice. Beyond the original issues in magnetism and critical phenomena physics that motivated its intensive exploration during the 20th century~\cite{Chaikin}, the Ising model and its extensions remain helpful to describe a wide range of phenomena in modern physics~\cite{Partridge2019,Maggi2021} and its applications. Applications go far beyond condensed matter physics, from the description of cell membranes~\cite{Cornet20} or animal skin patterning~\cite{Zakany2022}, to embryogenesis~\cite{Merle2019} and even to the dynamical description of geographical patterns~\cite{Ma2019}, to name a few. In most cases, using the Ising model supposes to have coarse-grained a more complex original system and one is in fact dealing with a mesoscopic model where microscopic degrees of freedom have been integrated out. In particular, simulating a mesoscopic, effective model enables one to save considerable amount of computational time by avoiding dealing with microscopic details. Then this Ising model can be coupled to any mesoscopic model of interest, for example a deformable membrane~\cite{Cornet20}.

Below the critical temperature, binary mixtures exhibit fluctuating boundaries between separated phases and determining the Ising parameter $J$ from the spectral density of their fluctuations follows a well-established procedure. When the parameter $J$ is just above its critical value, $J_c$, or equivalently said when the temperature $T$ is close enough to the critical one, $T_c$, the underlying lattice has a limited importance only and one recovers at large, macroscopic length-scales the features of a continuous theory, notably the isotropy of the original experimental system. For example, a droplet of the minority phase then has a globally roundish shape~\cite{Avron82,Shneidman01}, up to thermal fluctuations. The continuous limit will be extensively used in this work.

In the examples of applications given above, dynamical issues are generally at stake. An issue arises when one is interested in simulating the dynamics of the systems under study, because time-scales of the real system and of the numerical model must be precisely related. This is at the core of Kinetic Monte Carlo approaches relying on Monte Carlo local moves endowed with realistic dynamics~\cite{Newman}. Here we address this issue by calculating analytically the relaxation times of the fluctuation modes of a thermally activated boundary between phases, in different geometries of interest. We use the exact analogy between lattice gases and conserved-order magnetic systems and we follow the presentation of Bray~\cite{Bray94} (here in 2D). Our main contribution is to carefully take into account the renormalization of the different quantities entering the relaxation times close to $T_c$. In particular, the spontaneous magnetization and the magnetic susceptibility have non-trivial behaviors, described by universal critical exponents. However, thanks to the hyperscaling relation between these different critical exponents, we end with a remarkably simple expression of the relaxation times. At the end of the calculation, we use numerical simulations on the very simple Ising model to estimate the numerical prefactors, which are not universal and cannot be derived from renormalization group considerations. This answers the initial question, by relating explicitly the dynamics of the Ising model at the scale of local moves to the one of the interface fluctuation modes at large scales. The latter can in principle be measured either on experimental systems~\cite{esposito} or on Molecular Dynamics (MD) simulations~\cite{Marrink19}. Then one has access through our results to the real time-scale associated with local moves, to be implemented in the numerical mesoscopic models of interest.

\section{Stripe geometry, square and triangular lattices}

We consider an Ising model~\cite{Chaikin} on a square lattice of lattice spacing $a$. The Ising variables (also usually called ``spins'') are $s_i = \pm 1$ and the coupling is denoted by $J>0$. The critical temperature is~\cite{Baxter82} 
\begin{equation}
T_c = \frac{2}{\ln(1+\sqrt{2})}  \frac{J}{k_B} \simeq 2.27 \frac{J}{k_B}.
\label{Tc:sq}
\end{equation} 

\begin{figure}[h]
\centering
\includegraphics[height=5cm]{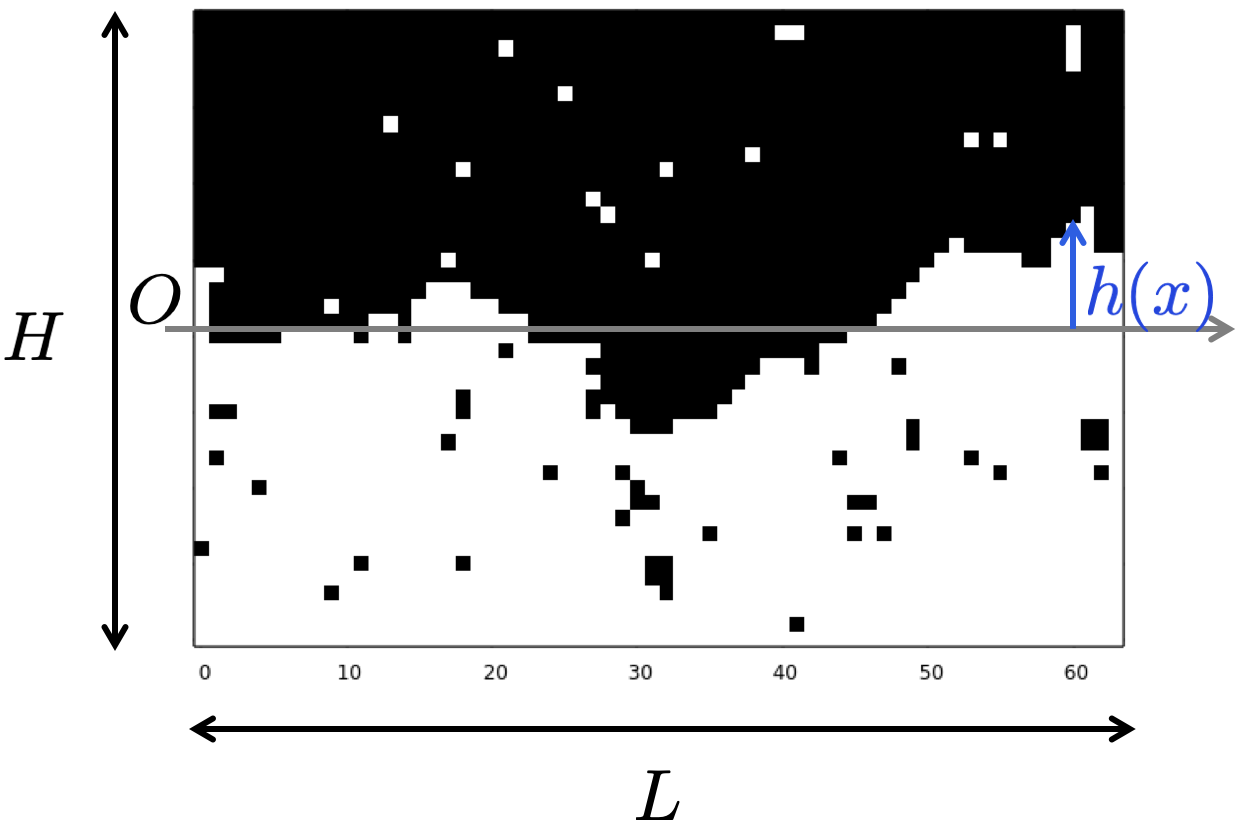}
\caption{Snapshot of Monte Carlo simulation of the Ising model in equilibrium on a $L=64$-long stripe at $T=1.8J/k_B$. White (resp. black) squares represent $s_i=-1$ (resp. $+1$) Ising variables. The simulation box height is $H=44$. The height-function $h(x)$ encoding the position of the interface between both phases is also shown in blue. Far from the interface, i.e. in the bulk, both phases coincide with the two equilibrium ones minimizing the free energy defined below. }
\label{stripe}
\end{figure}

The first system of interest is a stripe of length $L$ and height $H$, as shown in Figure~\ref{stripe}, at 0 total ``magnetization'' $\sum_i s_i$, so that there are exactly as many $+1$ and $-1$ Ising variables. The stripe height $H$ is sufficiently large so that the boundary hardly ever hits the stripe upper or lower sides ($H \gg \sqrt{L}$~\footnote{The height fluctuations $\sqrt{\langle h^2 \rangle} \propto \sqrt{L}$ for an interface with line tension $\lambda$, by using for example Eq.~\eqref{spec} and summing over modes $k$. Hence $\sqrt{\langle h^2 \rangle}  \ll H$. Consequently the interface hardly ever hits the upper or lower sides, as observed in practice in the simulations.}). Boundary conditions are set to $-1$ (resp. $+1$) on the lower (resp. upper) side, and are periodic between the vertical sides. Thus the height-function $h(x)$, giving the position of the interface (or domain wall) between both phases, is $L$-periodic. 

We denote by $\hat h_k(t)$ the Fourier coefficients of  
$h(x,t)$, $k \in \mathbb{Z}$:
\begin{equation}
\hat h_k(t) = \frac{1}{L} \int_0^L h(x,t) e^{-2i\pi k x/L} \, {\rm d}x
\end{equation}
The associated wave-length is $\Lambda = L/|k|$ for $k \neq 0$. We assume without loss of generality that $h(x)$ fluctuates around 0, that is to say $h_0 = \langle h(x,t) \rangle = 0$ by choosing properly the origin of the ordinates. 

We recall that in reality, the interface has a finite width in the $y$ direction, set by the correlation length $\xi$. At a distance much larger than $\xi$ from the interface, bulk phases coincide with the two equilibrium ones minimizing the free energy defined below. 

\subsection{Continuous limit}

We will adopt below a continuous field-theoretic approach where the discrete Ising variables $s_i = \pm 1$ become a continuous, coarse-grained real field $\phi(\mathbf{r})\in[-1,1]$, the local average of the $s_i$, governed by the Landau-Ginzburg  free-energy functional~\cite{Bray94,Chaikin}
\begin{equation}
F[\phi] = \int {\rm d}^2\mathbf{r} \left[  \frac12 (\nabla \phi)^2 + V(\phi)  \right]
\label{GL}
\end{equation}
In this relation, $V(\phi)=\frac{r}{2} \phi^2 + u \phi^4$ is a potential energy that, below the critical temperature $T_c$, favors the coexistence of two phases because $r<0$. At very low temperature, the two stable phases correspond to $\phi_+=1$ and $\phi_-=-1$, i.e. pure $+1$ and pure $-1$ phases. However, thermal agitation brings some disorder, so that  $\phi_+$ and $\phi_-$ get closer when $T$ grows and eventually become very close when reaching the critical temperature $T_c$. 

A very useful quantity below is the chemical potential defined by the functional derivative of $F[\phi]$:
\begin{equation}
\mu(\mathbf{r})=\frac{\delta F}{\delta \phi}
\label{mu:def}
\end{equation}
It is shown in Ref.~\cite{Bray94} that the bulk concentration relax much faster than the interface does, when the order parameter is conserved, so that the concentration field $\phi$ obeys Laplace's equation $\nabla^2 \phi=0$ in the bulk at the time-scales of interest. It ensues that $\mu$ depends linearly on $\phi$ in the bulk, and thus that $\mu$ also obeys Laplace's equation $\nabla^2 \mu=0$ there.

We wish to emphasize here that in principle the order parameter $\phi$ represents the magnetization field in the original Landau-Ginzburg theory, however it can also be considered as a local concentration after some basic algebraic manipulation: the field $\phi\in[-1,1]$, so that $c=(\phi+1)/2\in[0,1]$ is the local fraction of one species and $1-c$ the local fraction of the other species if the Ising model is seen as a lattice-gas, where $-1$ and $+1$ spins now represent two coexisting species (such as two atomic species in a binary alloy). Both quantities, magnetization and concentration, are equivalent. By extension, we shall generally use below the lattice-gas vocabulary, even though $\phi$ and $c$ do not exactly coincide. Notably, the chemical potential $\mu$ would correspond to the external magnetic field in the magnetism vocabulary.

The continuous limit is valid provided that $\xi/a \gg 1$. In other words, the temperature $T<T_c$ must be close enough to $T_c$. More precisely~\cite{Baxter82} 
\begin{equation}
\frac{\xi}a \simeq \frac{k_B T}{4J}\left(1-\frac{T}{T_c} \right)^{-1}
\label{xi:eq}
\end{equation}
for the 2D Ising model on a square lattice, which sets the regime of temperature for which the condition $\xi/a \gg 1$ is fulfilled (in general, the behavior of $\xi$ is governed by the critical exponent $\nu$~\cite{Chaikin}, equal to 1 in the present case). We also recall that a line tension, denoted by $\lambda$, is associated with the interface because it bears an energy cost proportional to its length. Thus $\lambda$ has the dimension of a force. As discussed below, it vanishes close to the critical point.  

\subsection{Evolution equation and relaxation times}

We consider a single mode $k>0$ of interface fluctuation, i.e. $h(x,t=0) = \alpha_0 \cos (qx)$ with $q= 2 k \pi/L=2 \pi/\Lambda $. We suppose that  $\alpha_0$ is small, i.e. much smaller than all relevant lenghscales apart from $a$. In particular, $\alpha_0 \ll \Lambda$, so that 
we work in the small gradient approximation, $|h'(x)| \ll 1$. 

We neglect the thermal noise in the interface dynamics, so that it will spontaneously return to its equilibrium position $h=0$, driven by the line tension $\lambda$: $h(x,t)$ is expected to be of the form $h(x,t) = \alpha_0 \cos (qx) e^{-t/\tau_q}$ for $t>0$~\footnote{Anticipating the exponential time dependence of $h(x,t)$ is not a prerequisite. Alternatively, one can just anticipate any factor $\ell(t)$ that would satisfy a first-order ODE at the end of the calculation.}. Langevin theory asserts that the relaxation time $\tau_q$ that we will determine below is equal to the decay time of the temporal correlation function of the Fourier coefficient that we shall measure in presence of thermal fluctuations and thus in numerical simulations.

Following Bray~\cite{Bray94}, section 2.4, one writes the evolution equation of $h(x,t) $ by calculating the interface velocity $v$ in the $y$ direction. We first need to compute the chemical potential $\mu(x,y,t)$. It satisfies Laplace's equation $\nabla^2 \mu=0$ far from the interface (see above), and it is subject to the Gibbs-Thomson-like boundary condition at the 1-dimensional interface (see Eq. (28) of Ref.~\cite{Bray94})]:
\begin{equation}
\mu(x,h(x),t) = - \frac{\lambda K(x)}{\Delta \phi}.
\label{bord}
\end{equation}
Here $\Delta \phi = \phi_+ - \phi_-$ is the difference of concentrations between both bulk phases (see below) and $K(x)$ is the local interface curvature. Introducing the vector $\mathbf{\hat g}=(g_x,g_y)$ normal to the interface~\cite{Bray94}, it writes
\begin{equation}
K(x) = \frac {{\rm d} g_x}{{\rm d} x} = -\frac{h''(x)}{[1+h'(x)^2]^{3/2}} \simeq -h''(x) = q^2 h(x).
\end{equation}
In addition, since $\alpha_0$ is assumed to be small, we can set $h(x)=0$ in the left-hand side of Eq.~\eqref{bord}, which now reads
\begin{equation}
\mu(x,0,t) = - \frac{\lambda q^2 h(x,t)}{\Delta \phi}.
\label{bord2}
\end{equation}
We naturally assume $\mu$ to be of the form $\mu(x,y,t)=\cos(qx) f(y) e^{-t/\tau_q}$ in the bulk, so that Laplace's equation leads to 
\begin{equation}
f''(y)-q^2f(y)=0. 
\end{equation}
Since $\mu$ cannot diverge at $\pm \infty$, the only physical solution is $f(y)=A e^{-q|y|}$. 

With the boundary condition, one eventually gets the chemical potential in the whole plane 
\begin{equation}
\mu(x,y,t) = - \frac{\lambda  \alpha_0 }{\Delta \phi} \, q^2 \cos(qx) e^{-q|y|} e^{-t/\tau_q}. 
\label{mu:full}
\end{equation}

We can now infer the interface velocity $v$ from $\mu$ by using the Eq.~(29) of Ref.~\cite{Bray94}, $v \Delta \phi = - \Gamma \left[\frac{\partial \mu}{\partial y} \right]_{-\epsilon}^{\epsilon}$, where the square brackets indicate the discontinuity across the interface. Indeed, the interface moves because of the imbalance between the currents flowing into and out of it, themselves proportional to the gradient of $\mu$ perpendicularly to the interface, $\frac{\partial \mu}{\partial y}$. Here, we have re-introduced a transport coefficient $\Gamma$ that is implicit in Bray, being ``adsorbed into the time scale''~\cite{Bray94}; $\Gamma^{-1}$ is homogeneous to a drag coefficient per unit area, arising from the continuity equation [below Bray's Eq.~(3), noted $\lambda$ therein]. We have also identified Bray's normal coordinate $g$ with $y$ in the small-gradient approximation. It follows that 
\begin{equation}
v(x,t) = - \frac{2 \Gamma \lambda  \alpha_0 }{(\Delta \phi)^2} \, q^3 \cos(qx) e^{-t/\tau_q} = - \frac{2\Gamma\lambda}{(\Delta \phi)^2} \, q^3 h(x,t)
\label{velocity}
\end{equation}
because we have anticipated that $h(x,t) = \alpha_0 \cos(qx) e^{-t/\tau_q}$. 
Note the $q^3$ factor, coming from $q^2$ in $\mu$ and the derivative $\frac{\partial \mu}{\partial y}$. Now, by definition, $v = \frac{\partial h}{\partial t} = - \frac1{\tau_q} h(x,t)$. It follows that 
\begin{equation}
\tau_q = \frac{(\Delta \phi)^2}{2\Gamma\lambda q^3}.
\label{tauk}
\end{equation}
Far below $T_c$, $\phi_{\pm}=\pm 1$ so that  $\Delta \phi \simeq 2$, while it decreases to 0 when $T$ goes close to $T_c$ as discussed above. We shall return to this point below.

\subsection{Introduction of the bulk diffusion coefficient}

We relate now the transport coefficient $\Gamma$ to the bulk diffusion coefficient $D$ in each phase. We start from the evolution equation governing the time evolution of  the (conserved) order parameter $\phi$ (model B, or Cahn-Hilliard equation)~\cite{Bray94,Chaikin}:
\begin{equation}
\frac{\partial \phi}{\partial t} = \Gamma \nabla^2 \frac{\delta F}{\delta \phi} 
\label{Langevin}
\end{equation}
where we have also re-introduced the transport coefficient $\Gamma$ implicit in Bray. Here $F[\phi]$ is again the free-energy functional of Eq.~\eqref{GL} with $V(\phi)$ the potential energy of the Landau-Ginzburg  theory~\cite{Chaikin}. In the bulk, if we now assume small perturbations of $\phi$ close to $\phi_+$ (or equivalently $\phi_-$) so that we can write $\phi = \phi_+ + \tilde \phi$, Eq.~\eqref{Langevin} becomes the diffusion equation [see below Bray's Eq.~(21)]:
\begin{equation}
\frac{\partial \tilde \phi}{\partial t} = \Gamma V''(\phi_+) \nabla^2 \tilde \phi
\label{Langevin2}
\end{equation}
Below $T_c$, $V(\phi)$ has two minima $\phi=\phi_\pm$. We assume for simplicity that $\phi_-=-\phi_+$, $V(\phi_+)=V(\phi_-)$ and $V''(\phi_+)=V''(\phi_-)$ because both bulk phases play the same role~\cite{Bray94}. Hence one can identify $\Gamma V''(\phi_+)$ with the diffusion coefficient in the bulk $D$, so that Eq.~\eqref{tauk} becomes
\begin{equation}
\tau_q = \frac{V''(\phi_+) (\Delta \phi)^2}{2 D \lambda q^3}
\label{tau:q}
\end{equation}
The line tension $\lambda$ being the driving force of the interface relaxation, $\tau_q$ naturally appears to be inversely proportional to $\lambda$ in this expression. It directly ensues from Eq.~\eqref{Langevin}, even though $\lambda$ is not the only factor to depend on $T$ in this relation as we shall see it now. 

\subsection{Vicinity of the critical temperature}

Now we determine how the factor $V''(\phi_+) (\Delta \phi)^2$ depends on the temperature close to $T_c$. We introduce the reduced temperature $\theta = T/T_c<1$. Firstly, $\Delta \phi$ is twice the bulk concentration close to $T_c$. Thus $\Delta \phi \propto (1-\theta)^\beta$, where we have used the definition of the critical exponent $\beta$~\cite{Chaikin}. Secondly, in the bulk, $\phi$ is uniform, thus the chemical potential in Eq.~\eqref{mu:def} is also uniform and reads $\mu=\delta F/\delta \phi = V'(\phi) - \nabla^2 \phi$. As already discussed, the last term is negligible in the bulk because $\phi$ obey's Laplace's equation at the time-scales of interest~\cite{Bray94} and 
\begin{eqnarray}
V''(\phi_+) & = & \frac{{\rm d}\mu}{{\rm d} \phi} (\phi=\phi_+) \\
& = &\frac1{\frac{{\rm d}\phi}{{\rm d} \mu} (\mu=0)} \\
& = & \frac{1}{\chi(\mu=0)} \\
& \propto& (1-\theta)^\gamma
\label{gamma}
\end{eqnarray}
where $\chi$ is the (magnetic) susceptibility and $\gamma$ the associated critical exponent. Indeed, using the previously mentioned analogy between the chemical potential $\mu$ (resp. the concentration $\phi$) and an external magnetic field, $h$ (resp. magnetization, $m$), we used  the definition of the magnetic susceptibility $\chi = \left(\frac{\partial m}{\partial h}\right)_T(h=0)$~\cite{Baxter82}. To go from the second to the third line above, we assumed that $\mu(\phi)$ is locally bijective in the neighborhood of $\phi_+$ and we used that $\mu(\phi_+)=0$ since $\phi = \phi_+$ (or  $\phi_-$) at  vanishing $\mu$. It follows that $V''(\phi_+) (\Delta \phi)^2 \propto (1 - \theta)^{2 \beta + \gamma} = (1 - \theta)^2$ owing to the hyperscaling relation between critical exponents~\cite{Chaikin} $2 \beta + \gamma = d \nu$, in dimension $d=2$ where $\nu=1$ for the Ising model~\cite{Baxter82}. The final behavior of $V''(\phi_+) (\Delta \phi)^2$ is remarkably simple. 

Beyond this scaling law, we are interested in the prefactors: $\Delta \phi$ has no dimension and $V''$ is proportional to $J/a^2$. Indeed, $J \propto k_BT_c$ is the only energy scale in the problem close to $T_c$ and there is one Ising variable per elementary square of area $a^2$ (see Ref.~\cite{Cornet20} for details). Hence 
\begin{equation}
\tau(\Lambda) = {\rm Const} \frac{J}{a^2 D \lambda} \Lambda^3  (1 - \theta)^2
\end{equation}
where we have written the relaxation time in function of the wavelength $\Lambda=2\pi/q$ and ${\rm Const}$ is a numerical constant. Contrary to the critical exponents, this prefactor is not universal, it depends on the microscopic details such as the underlying lattice. 

Furthermore $\lambda =4(1-\theta)J/a$ at first order in $1-\theta$~\cite{Avron82}, in other words, $1-\theta = \frac14 \lambda a /J$. Injecting this relation in the expression of $\tau(\Lambda)$, we get the simpler alternative expression
\begin{equation}
\tau(\Lambda) = A \frac{\lambda }{D J} \Lambda^3 
\label{tau:final}
\end{equation}
where $A$ is a new numerical constant, again non-universal. This is our main result : once the constant $A$ has been bench-marked on well-defined numerical systems (see below), the measure of $\tau(\Lambda)$ on experimental systems or MD simulations enables one to infer the bulk diffusion coefficient $D$, in real time units. 

This dependence of the time-scale on the cube of the length-scale, $\tau(\Lambda) \propto \Lambda^3$, is reminiscent of the law 
governing the coarsening of the same Ising model with conserved order-parameter after a quench from high to low temperature~\cite{Bray94}. In the latter case, the relaxation time of a length-scale $L$ also grows proportionally to $L^3$. Both mechanisms are closely related.

\medskip

\noindent {\em Remark 1}: In principle, the diffusion coefficient $D$ also depends on the temperature $T$. First of all, quantifying how (out-of-equilibrium) concentration fluctuations relax, as expressed by Eq.~\eqref{Langevin2}, $D$ is a \textit{cooperative} diffusion coefficient (also called \textit{mutual} or  \textit{collective} or \textit{gradient} diffusion coefficient), to be contrasted with the \textit{self}-diffusion coefficient describing the evolution of a single tagged particle~\cite{Chaikin,Scalettar88}. Note that both coincide in the limit of small density fluctuations. Critical phenomena theory states that $D$ goes to zero close to the critical point, which is related to critical slowing down. This is characterized by a dynamic critical exponent $z$ relating the correlation length $\xi$ and the correlation time $\tau$ through $\tau \sim \xi^z$ close to criticality. 
The diffusion coefficient then behaves like $D \sim \xi^2/\tau$ close to $T_c$~\cite{Honerkamp12}, i.e. (see also Ref.~\cite{Achiam80}):
\begin{equation}
D \sim \xi^{2-z} \propto (1-\theta)^{z'}
\end{equation}
with $z'=-\nu(2-z) = -(2-z)$ in the present case according to Eq.~\eqref{xi:eq}. If we specialize this result to the present Ising model with conserved order parameter (model B), a commonly accepted value is $z=4-\eta$ where $\eta$ is another critical exponent equal to $1/4$ in the present case~\cite{Chaikin}, leading to $z'=7/4$ (see Ref.~\cite{Honerkamp12} for recent numerical verification).  

In practice, however, when going away from the critical point, $D$ goes to a finite value $D_0$. Then the inverse diffusion coefficient can be interpolated by
\begin{equation}
D^{-1} \simeq  D_0^{-1} + C (1-\theta)^{-z'}
\label{D:renorm}
\end{equation}
as proposed in Ref.~\cite{Veatch08}, where $C$ is a model-dependent parameter that can be measured, from simulations or experiments. From now, we assume that $D_0$ dominates rapidly when one goes away from $T_c$, in particular for the values $\theta$ studied here, as observed experimentally in Ref.~\cite{Veatch08}, and we write $D \simeq D_0$. 

\medskip

\noindent {\em Remark 2}: In the Ising model of interest here, on a lattice with lattice spacing $a$, let us denote by $\delta t$ the (simulation) time step associated with local flips. Then $D=D_0=a^2/(4\delta t)$ for a freely bulk-diffusing single spin in a sea of opposite spins, i.e. at low enough density fluctuations.

Also using that $\lambda =4(1-\theta)J/a$, we can then also write 
\begin{equation}
\frac{\tau(\Lambda)}{\delta t} = A' (1-\theta) \left(\frac{\Lambda}{a}\right)^3 
\label{tau:final2}
\end{equation}
One expects $A$ and $A'=16A$ to be slowly varying functions of $\theta$ close to $T_c$, since the singularities are captured by the critical exponents.

\section{Monte Carlo simulation results in stripe geometry}

We begin with simulations on the square lattice with Kawasaki dynamics, at conserved order parameter~\cite{Newman}. In practice, one must be as close as possible to $T_c$ to use the scaling relations, as well as to use the continuous approach described above; however, close to $T_c$, the small-gradient approximation fails to describe the interface. A compromise must be found and we focus on the temperature range $\theta=0.6$ to $0.8$.  A simulation snapshot is given in Figure~\ref{stripe} at $\theta = 0.79$, i.e. $T=1.8J/k_B$. We have focused here on a relatively small system, $L=64$ and $H=44$, in order to have good statistical sampling. However, finite-size effects can affect the measurements of some thermodynamic quantities, such as the apparent localization of the critical temperature, or the apparent critical exponents. This has been thoroughly explored in the past~\cite{Newman}. Here, the principal limitation comes from the fact that the accessible wave-lengths $L/k$, $k$ integer, must be much larger than 1, so that the continuous limit remains meaningful. To ascertain that our numerical findings are not affected by such effects, we have performed a longer run on one larger $128\times 65$ system, as detailed below. Simulation durations were chosen so that more than 1000 statistically independent samples were computed for each condition (500 ones for $L=128$). This number was obtained by taking into account the slowest relaxation mode ($k=1$) of the phase boundary.



\subsection{Line tension}

We first extract the values of the line tension $\lambda$ and check their consistence w.r.t. the theoretical predictions. Owing to the Equipartition Theorem in thermodynamic equilibrium, the fluctuation spectrum depends on the line tension as~\footnote{This relation ensues from the Equipartition Theorem~\cite{Chaikin} after writing in the Fourier space the interface energy $\frac{\lambda}{2} \int_0^L | h'(x)|^2 {\rm d}x$  in the small-gradient approximation  (see for example Ref.~\cite{esposito} for the calculation in a similar context).}:
\begin{equation}
\langle | \hat h_k |^2 \rangle = \frac{L\, k_BT}{4 \pi^2 \lambda k^2}
\label{spec}
\end{equation}
In practice, we regularly measure the height-function $h(x,t)$ as follows. Confusion between the interface and small bubbles must be avoided (see Fig.~\ref{stripe}), all the more so as such bubbles becomes more probable when getting close to the transition. For each abscissa $x\in [1,L]$, starting from the bottom of the simulation box where $-1$ spins are the majority, we progress upward until we encounter a series of four consecutive $+1$ spins or more, or alternatively we reach the upper side (which is highly improbable because $H \gg \sqrt{L}$). This numerically defines $h(x)$, in units of the lattice spacing $a$. The choice of four consecutive spins above comes from the fact that for the temperatures studied here, we have observed bubbles of this size in the bulk phases to be very rare (see the figure). We have checked by visual inspection that this procedure faithfully captures the domain boundary for the range of temperature studied in this work.

Then we compute the Fourier coefficients $\hat h_k(t)$ with a FFT routine, and then average  $|\hat h_k|^2$ over simulation time. The values given below are obtained by fitting the three first modes, $k=1$ to 3. In table~\ref{tensionligne}, these numerical values are compared to the (asymptotically exact) theoretical ones close to $T_c$, i.e. $\lambda =4(1-\theta)J/a$. The agreement is good in spite of the diverse approximations used, such as the small gradient approximation or the above asymptotic behavior of $\lambda$ close to the critical point.

\begin{table}[h]
\centering
\begin{tabular}{c|c|c|c}
~$T/T_c$~ & ~$\lambda_{\rm theo}$~ & $\lambda_{{\rm num,64}}$ & $\lambda_{{\rm num,128}}$  \\
\hline
0.61 & 1.53 & $1.49 \pm 0.11 $ &   \\
0.70 & 1.18  & $1.16 \pm 0.05 $ & $1.14 \pm 0.07 $ \\
0.79 & 0.83 & $0.79 \pm 0.03 $ &   \\
\end{tabular}
\caption{Theoretically predicted and numerically measured line tensions for $L=64$ and 128-long stripes (square lattice) in function of the reduced temperature $\theta=T/T_c$, and in units of $J/a$. The duration of each simulation is $10^{12}$ and $12 \times 10^{12}$ for $L=128$) Monte Carlo steps. Error bars are 68\% confidence intervals. }
\label{tensionligne}
\end{table}

\medskip

\noindent {\em Remark}: To check the validity of the small gradient approximation, one can estimate the order of magnitude of the gradient $\partial h/\partial x$. Owing to the discrete version of Parseval's identity, the average value of $|\partial h/\partial x|^2$ is computed as follows
\begin{eqnarray}
\Big\langle \frac1{L} \int_0^L \left|\frac{\partial h}{\partial x}\right|^2 {\rm d}x \Big\rangle & = & \sum_k q_k^2  \langle |\hat h_k|^2 \rangle \\
 & = & \sum_{k=0}^{L/a-1}  \left(  \frac{2 k \pi}{L} \right)^2  \frac{L\, k_BT}{4 \pi^2 \lambda k^2} \\
 & = & \frac{k_BT}{a \lambda}
\end{eqnarray} 
All modes contribute equally. Here owing to table~\ref{tensionligne},  $\lambda \approx J/a = \frac{\ln(1+\sqrt{2})}{2} k_B T_c /a$. Thus $  \frac{k_BT}{a \lambda} \approx  \frac2{\ln(1+\sqrt{2})} \frac{T}{T_c}$. It follows that $\partial h/\partial x$ is already of order 1 with the temperatures studied here and one cannot reasonably go closer to $T_c$ where $\lambda$ would tend to zero and the gradient diverge.

\subsection{Relaxation times}

\begin{figure}[t]
\centering
\includegraphics[height=5cm]{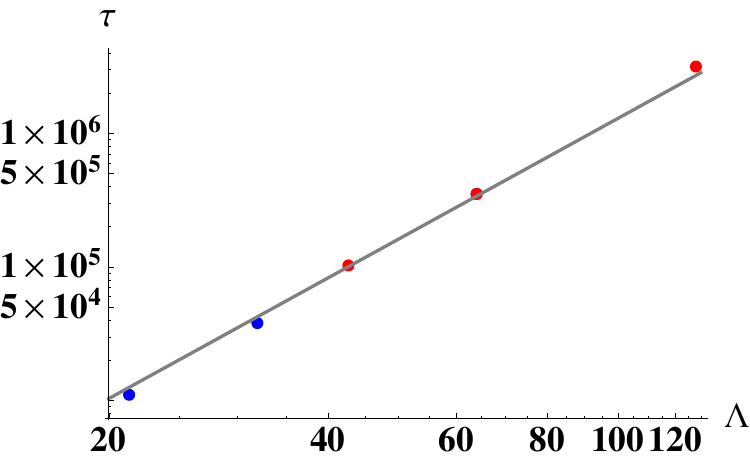}
\caption{Relaxation times (in units of $\delta t$) of the 3 slowest modes ($k=1$ to 3) in function of the wave-length $\Lambda=L/k$ (in units of $a$), for  the reduced temperature $\theta=0.70$ and two system sizes $L=64$ (blue dots) and $L=128$ (red dots). The two points at $\Lambda=64$ are almost superimposed. Log-log coordinates. The continuous line has slope 3 for comparison with the expected behavior of Eq.~\eqref{tau:final}.
\label{tau:fig}}
\end{figure}

Once line tensions have been determined, we can use Eq.~\eqref{tau:final} to calculate the value of the numerical constant $A$. In the simulation units, $D=1/4$, implicitly in units of $a^2$ per Monte Carlo sweep of duration $\delta t$~\cite{Newman}. In practice, for a fixed mode $k$ (or $q = 2k\pi/L$), to compute $\tau_k$, the coefficients $\hat h_k(t)$ are Fourier-transformed with respect to time (again with the help of FFT), the new Fourier coefficients being denoted by $\tilde h_k(\omega)$. We are interested in the correlation function $\hat C_k(s)= \langle \hat h_k(t) \hat h_k^*(t+s) \rangle$, where the star denotes the complex conjugate. Owing to Wiener-Khinchin's theorem one finally gets
\begin{equation}
\hat C_k(s) \propto {\rm FT}^{-1}\left[ \left| \tilde h_k(\omega) \right|^2  \right](s)  
\end{equation}
where ${\rm FT}^{-1}$ is the inverse Fourier transform w.r.t. time, from which we can extract the relaxation time $\tau_k$ of each mode $k$ by fitting $\hat C_k(s)$ in normal-log coordinates on its linear regime~\cite{Camley2010}. Fig.~\ref{tau:fig} provides examples of measured relaxation times, and illustrates that $\tau(\Lambda) \propto \Lambda^3$.

Using Eq.~\eqref{tau:final}, and averaging the measured $\tau(\Lambda)/\Lambda^3$ over the 3 slowest modes, we eventually find $A\simeq 0.26$ for $\theta=0.61$, $A\simeq 0.26$ for $\theta=0.70$ (and $A\simeq 0.29$ if $L=128$ at this temperature), and $A\simeq 0.32$ for $\theta=0.79$ on the square lattice. The relative lack of accuracy on the value of $A$ principally comes from two antagonistic constraints: the deviation from the small gradient approximation increases when $T$ gets too close to $T_c$ where $\lambda$ vanishes, and by contrast, the first-order expansion of Eq.~\eqref{tau:final} is valid only close to $T_c$. 

\subsection{Triangular lattice}

We now confront our analytical results to numerical ones on an alternative 2D lattice. We adapt the above simulation procedure to the triangular lattice. In addition to the N, S, W and E edges of the square lattice, we add NW and SE ones so that each vertex in the bulk has 6 now nearest neighbors. The boundary conditions are unchanged as compared to the previous case. 

The critical temperature becomes \cite{Baxter82}
\begin{equation}
T_c = \frac{4}{\ln 3}  \frac{J}{k_B} \simeq 3.64 \frac{J}{k_B}.
\label{Tc:tri}
\end{equation} 
Some care must be taken of the definition of the height function with the chosen boundary conditions. Indeed, we detect the boundary along the sequence of spins  parallel to the $y$ axis, at fixed abscissa. The so-obtained height must then be rescaled by a geometrical factor $\sqrt{3}/2$ because the actual lattice unit cell is an equilateral triangle of height $\sqrt{3}a/2$ along the $y$ axis.  In table~\ref{tensionligne2}, numerical values of the so-obtained line tension $\lambda$ are again compared to the theoretical ones close to $T_c$, $\lambda =4\sqrt{3}(1-\theta)J/a$ on a triangular lattice~\cite{Shneidman01}, after following the same process as above. The agreement is almost as good as in the square lattice case.

\begin{table}[h]
\centering
\begin{tabular}{c|c|c}
~$T/T_c$~ & ~$\lambda_{\rm theo}$~& ~$\lambda_{\rm num}$~  \\
\hline
0.71 & 1.98 & $2.10 \pm 0.08$  \\
0.77 & 1.60 & $1.73 \pm 0.07$  
\end{tabular}
\caption{Theoretically predicted and numerically measured  line tensions for $L=64$-long stripes (triangular lattice) in function of the reduced temperature $\theta=T/T_c$, in units of $J/a$. The duration of  each simulation is $10^{12}$ Monte Carlo steps. Error bars are 68\% confidence intervals.}
\label{tensionligne2}
\end{table}

Furthermore, using Eq.~\eqref{tau:final}, and again averaging the measured $\tau(\Lambda)/\Lambda^3$ over the 3 slowest modes, we find on the triangular lattice $A\simeq 0.10$ for $\theta=0.71$ and $A\simeq 0.12$ for $\theta=0.77$. In this case, Eq.~\eqref{tau:final2} still holds with $A'=16 \sqrt 3 A$. The value of $A$ differs from the square-lattice one owing to the non-universal character of this prefactor.

\section{Circular geometry, polar coordinates}

We now consider a quasi-circular droplet of $+1$ spins in a sea of $-1$ spins. The droplet radius is $R_0\gg a$ if it were perfectly circular, i.e. its (conserved) area is $\pi R_0^2$. We recall that for temperatures close enough to $T_c$, the isotropy of the system is restored at large length-scales, independently of the lattice symmetries. We again perturb the circular droplet by considering only the mode $k$.  Taking into account conservation of the order parameter is slightly more complex than in the stripe geometry above. If we set the origin at the center of the original circle, the boundary shape in polar coordinates reads $r(\theta)=R_0(1+u_0+u_k \cos(k\theta))=r_0+\rho_0 \cos(k\theta)$, considering again the mode $k$. Here $r_0=R_0(1+u_0)$ and $\rho_0=R_0 u_k \ll r_0$. Indeed  one must keep the mode $k=0$, $u_0$, because conservation of domain area imposes a relationship between the different modes $u_k$~\cite{esposito}. It follows that $r''(\theta)=k^2[r_0-r(\theta)]$. 

In polar coordinates, the curvature is given that
\begin{equation}\label{K}
    K= \frac{r(\theta)^2+2r'(\theta)^2 -r(\theta)r''(\theta)}{[r(\theta)^2+r'(\theta)^2]^{3/2}}
    \simeq \frac{1}{r(\theta)}-\frac{r''(\theta)}{r^2(\theta)}=\frac{1}{r(\theta)}(1+k^2)-\frac{k^2r_0}{r^2(\theta)}
\end{equation}
in the small gradient approximation ($\rho_0 \ll r_0$) where $r'(\theta)$ and $r''(\theta)\ll r(\theta)$.  At ordre 1, one also gets
\begin{equation}
    \frac{1}{r(\theta)}=\frac{1}{r_0}-\frac{\rho_0}{r_0^2}\cos(k\theta)
\end{equation}
and $K$ becomes
\begin{equation}
    K=\frac{1}{r_0}\Big[1+(k^2-1)\frac{\rho_0}{r_0}\cos(k\theta)\Big]
\end{equation}

To determine the chimical potential $\mu$ in the whole plane, we use again the boundary condition
$\mu(\theta, r(\theta),t)
\simeq \mu(\theta, r_0,t) =-\frac{\lambda K(\theta)}{\Delta \phi}$. In polar coordinates, the Laplace equation $\nabla^2 \mu=0$ reads
\begin{equation}
    \frac{\partial^2\mu}{\partial r^2}+\frac{1}{r}\frac{\partial \mu}{\partial r}+\frac{1}{r^2}\frac{\partial^2 \mu}{\partial \theta^2}=0
\end{equation}

We first determine $\mu$ {\em outside} the domain, i.e. for $r>r_0$. We look for a solution of the form
$\mu(\theta,r,t)=\mu_0(r) +
f(r)\cos(k\theta)e^{-t/\tau_k}$, where $\mu_0(r)=-\frac{\lambda}{\Delta \phi}\frac{\ln r}{r_0 \ln r_0}$is the solution in 2D in absence of boundary fluctuations. It follows that
\begin{equation}\label{laplace_pol}
    r^2f''(r)+r f'(r)-k^2f(r)=0
\end{equation}
Looking for power a law solution, $f(r)=B r^\alpha$, and injecting it in Eq.~(\ref{laplace_pol}), we get 
\begin{equation}
    \alpha(\alpha-1)B r^\alpha+\alpha B r^\alpha-k^2B r^\alpha= (\alpha^2-k^2)B r^\alpha=0
\end{equation}
There are two independant solutions, $\alpha=\pm|k|$ for $|k|>1$, and we choose $\alpha=-|k|$, the only solution remaining finite at infinity. The constant $B$ is determined through the boundary condition $\mu(r_0,\theta)=-\frac{\lambda}{r_0 \Delta \phi}-(k^2-1)\frac{\sigma_0 \rho_0}{r_0^2} \cos(k \theta)$, we find $B={\rm Const.} \ r_0^{|k|}$. Finally, for $r>r_0$, 
\begin{equation}
    \mu(\theta, r,t)=\frac{\lambda}{\Delta \phi}
    \Big[-\frac{1}{r_0}\frac{\ln r}{\ln r_0}+\frac{\rho_0}{r_0^2}(k^2-1)\left(\frac{r_0}{r}\right)^{|k|}\cos(k\theta)\Big]e^{-t/\tau_k}
\end{equation}

{\em Inside} the domain, i.e. for $r<r_0$, the trivial solution in absence of fluctuations would be $\mu_0={\rm Const.} = -\frac{\lambda}{r_0 \Delta \phi}$. This time, we look for a solution being regular at the origin $O$, so that we only keep the solution $\alpha=+|k|$:
\begin{equation}
    \mu(\theta, r,t)=\frac{\lambda}{\Delta \phi}
    \Big[-\frac{1}{r_0}+\frac{\rho_0}{r_0^2}(k^2-1)\left(\frac{r}{r_0}\right)^{|k|}\cos(k\theta)\Big]e^{-t/\tau_k}
\end{equation}

We look for the discontinuity at $r_0$:
\begin{equation}
    \begin{cases}
        \vspace{5mm}
        r>r_0: \frac{\partial \mu}{\partial r}(r_0+\epsilon)=\Big[-\frac{\lambda}{r_0^2 \ln r_0 \Delta \phi}-\frac{\lambda \rho_0}{r_0^3 \Delta \phi}(|k|^3-|k|)\cos(k\theta)\Big]e^{-t/\tau_k}\\
        r<r_0: \frac{\partial \mu}{\partial r}(r_0-\epsilon)=\frac{\lambda \rho_0}{r_0^3 \Delta \phi}(|k|^3-|k|)\cos(k\theta) e^{-t/\tau_k}
    \end{cases}
\end{equation}
where $\epsilon>0$ is vanishingly small. The interface velocity $v(\theta, t)$ is now given by~\cite{Bray94} $v\Delta\phi=-\Gamma \Big[\frac{\partial \mu}{\partial r}\Big]^{r_0+\epsilon}_{r_0-\epsilon}$. We are only interested by the contribution of $v$ on the relaxation of mode $k$. Indeed, there is also a contribution acting on $r_0$ and making it  go to 0 at large times. In standard coarsening theory, this expresses the evaporation of small domains to the benefit of largest ones, elsewhere in the system (Ostwald ripening)~\cite{Bray94}. Here, we are not interested in this very long time-scale process because we consider a single isolated domain on shorter time-scales.  As $u_0 = \mathcal{O}(u_k^2)$~\cite{esposito}, there is no coupling between modes 0 and $k$ at order 1. Thus
\begin{equation}
    v(\theta,t)= - 2\frac{\lambda \Gamma \rho_0}{r_0^3 (\Delta\phi)^2}
    (|k|^3-|k|) \cos(k\theta) e^{-t/\tau_k}
    =\frac{\partial r}{\partial t} =    -\frac{1}{\tau_k}r(\theta,t)
\end{equation}
leading to
\begin{equation}
    \tau_k=\frac{r_0^3 (\Delta\phi)^2}{2\Gamma \lambda}  \frac1{|k|^3-|k|}
\label{tau:lambda:circ}
\end{equation}
Note that $k=1$ is a soft mode in polar geometry, corresponding to the translation of the domain center. Its relaxation time is thus irrelevant in this geometry. 

The connection with the above Cartesian geometry can be done by identifying $q=k/r_0=2\pi k/L$, because the unperturbed interface length is $L=2 \pi r_0$. At large $k$, $|k|^3-|k| \simeq |k|^3$
and we recover
\begin{equation}
\tau_k \simeq \frac{(\Delta\phi)^2}{2\Gamma \lambda q^3},
\end{equation}
as in Cartesian geometry, see Eq.~\eqref{tauk}, because the interface is locally flat at the scale of the wave-length $\Lambda \ll R_0$. Also note that the chemical potentials in polar and Cartesian geometries are equal close to the interface. Indeed, let us for example consider the neighborhood of the point {$(0,r_0)$ of the interface and write $(x,y)=(0,r_0+Y)$, where $Y$ is the distance to the l'interface. If $r>r_0$, i.e. $Y>0$, $(r_0/r)^k=\exp[-k \ln(1+Y/r_0)]\simeq \exp(-qY)$ provided that we identify $q=k/r_0=2\pi k/L$, with $L \simeq 2\pi r_0$. In the same way $(r/r_0)^k \simeq \exp(qY)$ close to the interface if $r<r_0$, i.e. $Y<0$.

To conclude, since the large $k$ behavior is the same for stripe and circular geometries, it follows that the numerical prefactor $A$ determined in stripe geometry above, depending on the underlying lattice, can also be used here to relate numerically the relaxation times to the model parameters, as follows
\begin{equation}
\tau(\Lambda) = A \frac{\lambda }{D J} \frac{\Lambda^3}{1-\left(\frac{\Lambda}{2\pi r_0}\right)^2}
\label{tau:final:polar}
\end{equation}
This relation results from Eq.~\eqref{tau:lambda:circ}, by using the same calculation as above when deriving Eq.~\eqref{tau:final} from Eq.~\eqref{tauk}, but in circular geometry instead of stripe geometry.

\section{Discussion}

We come back to the initial issue raised in the introduction: How to extract the real time-scale associated with Kawasaki local moves from the measurement of macroscopic time-scales of boundary relaxation modes? Assume that we can localize and track experimentally or in MD numerical simulations the interface between two phases, either in stripe or in polar (droplet) geometry. To gain computational efficiency and address larger system sizes on longer timescales, the system can be advantageously simulated with the help of the coarse-grained Ising model on a square or a triangular lattice~(see, e.g., Ref.~\cite{Cornet20}) provided that (i) the Ising coupling $J$ and (ii) the time-step $\delta t$ associated with each Monte Carlo step are suitably chosen.  They significantly depend on the choice of the lattice spacing $a$~-- that must be chosen to be smaller than all length-scales of interest in the problem considered (except of course the molecular ones)~--. We propose the following protocol to properly determine $J$ and $\delta t$.
\begin{enumerate}
\item Measure the line tension $\lambda$ with the help of the interface fluctuation spectrum in equilibrium measured on real systems or MD simulations, as prescribed by Eq.~\eqref{spec}~\cite{esposito}.
\item $T$ is set by the experimental conditions, thus $J_c$ is given by $J_c = \frac{\ln(1+\sqrt{2})}2 k_B T$ (resp. $ \frac{\ln 3}4 k_B T$) on a square (resp. triangular) lattice. Then $J$ can be estimated close enough to $J_c$ by known exact expansions, $J-J_c \propto \lambda a$, or by exact results farther from $J_c$~\cite{Baxter82,Avron82,Shneidman01}. 
\item By measuring $\tau(\Lambda)$ on real systems or MD simulations, one estimates $D$ with our relation~\eqref{tau:final} and the numerical coefficient $A$ corresponding to the appropriate lattice.
\item From the knowledge of $D$ and the appropriate choice of $a$ (shorter than any relevant length-scale above the molecular ones), one eventually deduces the real time-step $\delta t$ corresponding to a Monte Carlo step through $\delta t = a^2/(4D)$ in two dimensions. 
\end{enumerate}

In the model-B context under consideration here, composition fluctuations dissipate through the diffusion of microscopic constituents. The present work can in principle be extended to contexts where additional hydrodynamic effects are taken into account. This is for example discussed into detail in Refs.~\cite{Camley2010,Camley2010B,Honerkamp12} in the context of biphasic lipid membranes, where the internal membrane dynamics can additionally be coupled to the 3D hydrodynamics of the surrounding solvent. Instead of the model B used here, one would need to appeal to the so-called models H or HC as discussed in these references. Relating large-scale dynamics to coarse-grained Ising-like model ones remains to be done in this more complex situation. 


\subsection{Acknowledgments} We are grateful to Manoel Manghi and Matthieu Chavent for fruitful discussions on this work.

\end{document}